\documentclass[aps,pre,twocolumn,showpacs]{revtex4}
\usepackage{epsfig}
\usepackage{amsmath}
\usepackage{amssymb} 
\usepackage{graphicx}
\usepackage{graphics}
\begin{document}
\date{\today}
\title{Self-organization of the MinE ring in subcellular Min oscillations}
\author{Julien Derr$^{1,2}$} 
\email{julien.derr@espci.org}
\author{Jason T. Hopper$^1$}
\author{Anirban Sain$^3$}
\author{Andrew D. Rutenberg$^1$}
\email{andrew.rutenberg@dal.ca}
\affiliation{$^1$Department of Physics and Atmospheric Science, Dalhousie University, Halifax, Nova Scotia B3H 3J5, Canada \\
$^2$ FAS Center for Systems Biology, Harvard University, Northwest Labs, 52 Oxford Street, Cambridge, MA 02138,              
USA \\
$^3$ Physics Department, Indian Institute of  Technology-Bombay, Powai 400076, India}

\date{\today}
\begin{abstract} 
We model the self-organization of the MinE ring that is observed during subcellular oscillations of the proteins MinD and MinE within the rod-shaped bacterium {\it Escherichia coli}.  With a steady-state approximation, we can study the MinE-ring generically -- apart from the other details of the Min oscillation. Rebinding of MinE to depolymerizing MinD filament tips controls MinE ring formation through a scaled cell shape parameter $\tilde{r}$.  We find two types of E-ring profiles near the filament tip: a strong plateau-like E-ring controlled by 1D diffusion of MinE along the bacterial length, or a weak cusp-like E-ring controlled by 3D diffusion near the filament tip. While the width of a strong E-ring depends on $\tilde{r}$, the
occupation fraction of MinE at the MinD filament tip is saturated and hence the depolymerization speed do not depend strongly on $\tilde{r}$. Conversely, for weak E-rings both $\tilde{r}$ and the MinE to MinD stoichiometry strongly control the tip occupation and hence the depolymerization speed. MinE rings {\em in vivo} are close to the threshold between weak and strong, and so MinD-filament depolymerization speed should be sensitive to cell shape, stoichiometry, and the MinE-rebinding rate.   We also find that the transient to MinE-ring formation is quite long in the appropriate open geometry for assays of ATPase activity {\it in vitro}, explaining the long delays of ATPase activity observed for smaller MinE concentrations in those assays without the need to invoke cooperative MinE activity.  
\end{abstract}
\maketitle

\section{Introduction}
The oscillation of the proteins MinD and MinE from pole to pole of individual cells of the bacterium {\it Escherichia coli} is used to localize cellular division to midcell \cite{reviews}. One cycle of the oscillation, lasting approximately one minute, starts with ATP-associated MinD binding to the bacterial inner membrane and polymerizing into helical filaments \cite{Shih2003, Hu2002, Suefuji2002} (see also \cite{DillonFilaments}). This occurs at alternating poles of the bacterium, with the MinD forming a polar ``cap''.  MinE is recruited to the membrane-bound MinD, where it forms a distinctive  ``E-ring''  \cite{Raskin1997,Hale2001,Fu2001} at the edge of the MinD cap by accumulating near the MinD filament tips \cite{Shih2003}.  Because the rate of hydrolysis and subsequent release of ATP-MinD is stimulated by MinE \cite{Hu2001, Hu2002, Suefuji2002}, the E-ring drives depolymerization of the MinD filament which allows the oscillation to proceed.  The depolymerization occurs with an approximately fixed E-ring width and speed along the cell axis \cite{Fu2001,Hale2001}, indicating an approximate steady-state during this part of the Min oscillation.  However, little is known about the mechanism of E-ring formation, its detailed structure, or how important it is for Min oscillations. Indeed, Min oscillations have been observed without prominent E-rings \cite{Shih2002}. 

Most models proposed for Min oscillation do not have explicit MinD filaments \cite{Howard2001,Howard2003,Huang2003,Meinhardt2001,Kruse2002}, though they do have E-rings.  Recently, several models of Min oscillations that include explicit MinD polymerization have been proposed \cite{Pavin2006,Tostevin2006,Drew2005,Cytrynbaum2007}, two of which display strong E-rings that track the tips of depolymerizing MinD filament caps with constant speed and width \cite{Cytrynbaum2007,Drew2005}.  In these models, E-rings are the result of MinE polymerization either orthogonal to \cite{Drew2005}, or along \cite{Cytrynbaum2007}, MinD filaments.   While MinD polymerization has been observed {\em in vitro} \cite{Hu2002,Suefuji2002}, there have been no reports of MinE polymerization in the experimental literature.  Indeed, the faint MinE ``zebra-stripes'' associated with the MinD zones adjacent to the MinE ring \cite{Hale2001,Fu2001,Shih2002} seem to imply sparse lateral binding of MinE to the body of MinD filaments -- not MinE polymerization.  

In this paper, with both stochastic 3D simulations, and a deterministic 1D model, we show that local (non-polymeric) rebinding of MinE released from depolymerizing MinD filament tips is sufficient for E-ring  formation.  We impose and characterize a dynamical steady-state of an E-ring on a depolymerizing semi-infinite MinD filament in order to address the approximate steady-state speed and width of the E-ring {\em in vivo} \cite{Hale2001,Fu2001}. We investigate the roles of spatial dimension, cell length, and radius, and of multiple MinD filaments and their helical pitch. We estimate the timescale of E-ring formation and obtain results consistent with the significant delay before ATPase activity seen with small MinE concentrations and large MinD membrane coverage {\em in vitro} \cite{Hu2002,Suefuji2002}.  Finally, we discuss how competition between the intrinsic and the MinE-stimulated ATPase activity of MinD controls the instability that leads to the initial formation of the E-ring from a uniformly decorated MinD filament. 

Qualitatively, we predict that the width of MinE-rings will increase as the MinD-filament depolymerization speed is increased through manipulation of cell shape, MinD to MinE stoichiometry, or mutations that affect the MinE binding rate to MinD.  Eventually, the depolymerization speed will saturate but the E-ring width can still grow. Conversely, as the depolymerization speed is decreased, MinE-rings will undergo a transition from a plateau-like ``strong'' E-ring to a cusp-like ``weak'' E-ring.   To our knowledge, systematic experimental studies of the E-ring width have not yet been done.  

\begin{figure} \begin{center} \includegraphics[width=\linewidth]{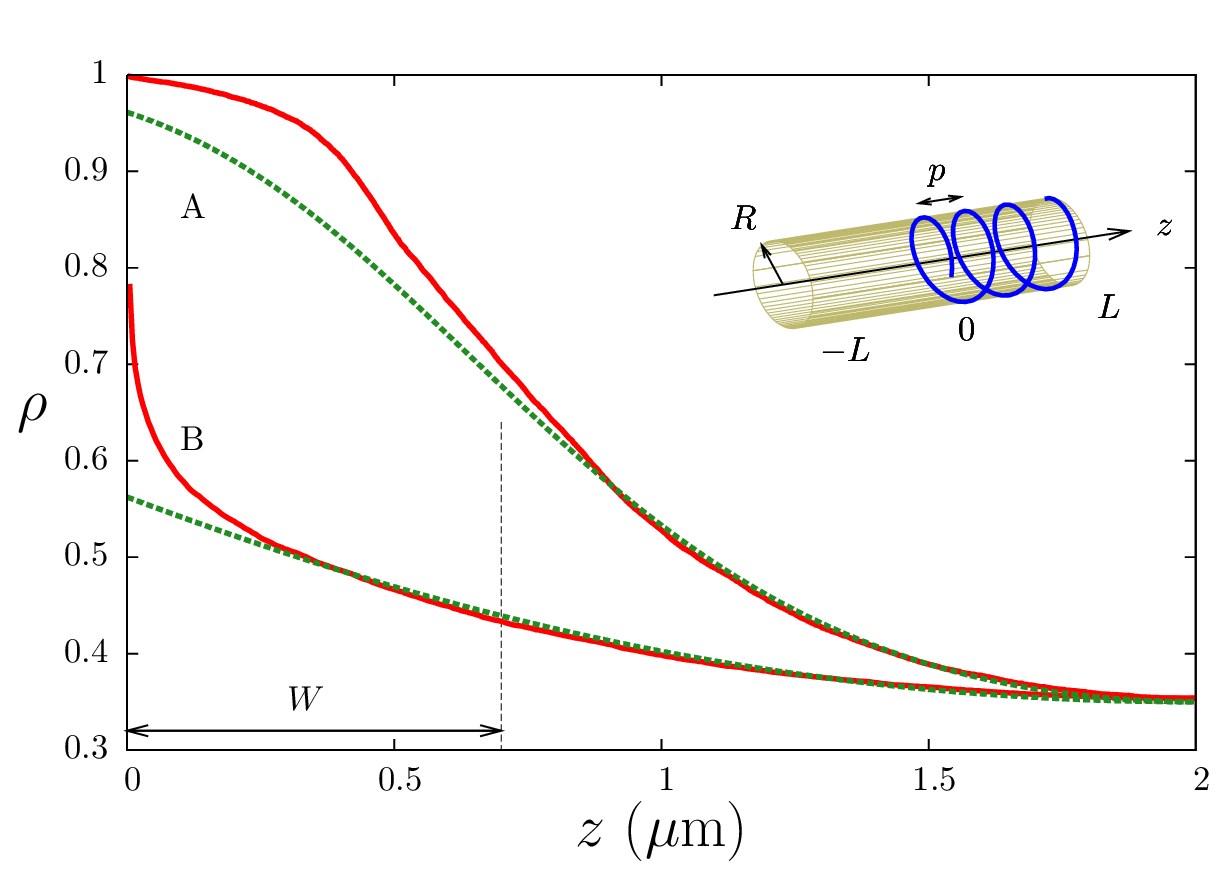}
\caption{Fractional occupancy of MinE on the MinD filament $\rho$ vs. distance along the bacterial axis $z$ for the 3D stochastic model (continuous lines) and the 1D model (dashed lines) for parameters typical of {\it E. coli}: $L=2~\mu$m, $R=0.5 ~\mu$m, and $\rho_0=0.35$. The MinE binding parameter was $\sigma_3=0.3 \mu m^3 /s$, while for the 1D model $f=0.06$ was used. One MinD filament supports either (A) a strong plateau-like E-ring for pitch $p=0.45~\mu$m or (B) a weak cusp-like E-ring for pitch $p=\infty$ (straight filament). The width $W$ of the strong E-ring, given by $\rho(W)=(1+\rho_0)/2$ is indicated. The inset illustrates the cylindrical geometry of the 3D model, showing the underlying helical MinD filament with its 
depolymerizing tip at $z=0$. The helical pitch $p$ is indicated.}
\label{fig:profiles} 
\end{center} \end{figure}

\section{E-ring Model}
As illustrated in the inset of Fig.~\ref{fig:profiles}, we represent the bacterial geometry as a cylinder of radius $R$ and length $2L$. In the right half ($0<z<L$), $n$ filaments of MinD are placed on the cylinder, each with the same helical pitch $p$ but with random (unbundled) helical phases. MinD filaments are composed of monomers of length $a_0$, each of which can bind one MinE.  We depolymerize MinD from filament tips at $z=0$, and any released MinE diffuses in the cylinder interior (cytoplasm) with a diffusion constant $D$.  Released MinE can bind to unoccupied MinD monomers; if not it is removed from the system at  $z=\pm L$. This open boundary condition represents the sinks for MinE provided by other MinD in the system.  Depolymerized MinD is removed from the system without further interaction, reflecting the nucleotide exchange needed before MinD rebinding is possible.  This dramatically simplifies our model, since we may then explicitly consider only MinE dynamics on an implicit MinD filament. Both the boundary conditions and the neglect of depolymerized MinD will be addressed again in the discussion.

In order to study a steady-state E-ring, we keep the filament tips centered at $z=0$ -- the ``tip-frame''.  
In the tip-frame, bound MinE move along MinD filaments at a constant depolymerization speed $v$ while new monomers of MinD are introduced at $z=L$ decorated with MinE with a constant probability $\rho_0$ (determined by the relative cellular amounts of MinE and MinD particles).  [Effectively we are studying semi-infinite MinD filaments under the approximation of uniform MinE binding for $z>L$.]  In the steady-state, the fraction of MinE released by the depolymerizing MinD filament tip that reach the absorbing boundaries will then be $\rho_0/\rho_{tip}$, where $\rho_{tip}$ is the fractional MinE occupation of the filament tip. The depolymerization speed $v$ 
can be determined self-consistently by $\rho_{tip}$, though we will see below that $v$ is small and can be practically ignored in terms of the E-ring structure.

\subsection{Stochastic 3D implementation}
The dimensionless parameter $\alpha_\ell \equiv v \ell/ D$, the fractional axial distance one MinE advects at speed $v$ while it diffuses a distance $\ell$, characterizes the importance of the depolymerization speed. Even with $\ell = 4 \mu m$, $v = 0.03 \mu m/s$ \cite{Hale2001,Fu2001}, and $D=10 \mu m^2/s$ \cite{Meacci2006},  $\alpha_\ell = 0.01$ is small and depolymerization is slow compared to diffusion. Accordingly, our stochastic 3D model quasiadiabatically follows each released MinE until it either rebinds or is removed from the system before allowing further depolymerization.  Each MinE diffuses by taking a randomly oriented step of fixed length $\delta$ every timestep $\Delta t$, where $D=\delta^2/(6 \Delta t)$. Diffusing MinE binds to a free MinD with probability $P_{stick}$ when it hits the bacterial membrane within a distance $r_{bind}$ of the MinD. We take $r_{bind}=a_0$.  This leads to an effective binding rate of $\sigma_{3} \rho_{3,local}$, where $\rho_{3,local}$ is the local bulk concentration of MinE and the bulk reaction rate $\sigma_{3}= 3 \pi D r_{bind}^2 P_{stick}/(2 \delta)$.  We take $\sigma_3= 0.3$~$\mu m^3/s$ (this is approximately the threshold between strong and weak E-rings given the cell geometry, see below). The steady state reached after successive depolymerization steps is independent of small $\delta$ if we vary $P_{stick}$ with $\delta$ to keep $\sigma_3$ constant.

\subsection{Analytic 1D treatment}
We also study a deterministic 1D model that exactly 
corresponds to the 3D stochastic model in the limit $R \ll a_0$. This enables us to explore the role of spatial dimension and stochastic effects in the E-ring, and also helps us to identify the combinations of parameters that control the E-ring structure.  Our 1D model tracks both the linear density of bound MinE (B) and of freely diffusing MinE  $(F)$:
\begin{eqnarray}
	\dot{B} -v B' & = & \sigma_1 F (B_{max}-B)  \text{ , for } z>0 \\
	\dot{F} -v F' & = & D F'' - \sigma_1 F (B_{max}-B)+vB(0) \delta(z), \label{EQN:unscaledF}
\end{eqnarray}
where the dots and primes indicate time and spatial derivatives, respectively. For $z<0$ there are no filaments so $B=0$.  For $z>0$, the linear density of potential binding sites (i.e. of MinD) is $B_{max} = 1/a$, and the 1D rebinding rate is  $\sigma_1$. The $v$ dependent terms on the left side of the equations represent advection of bound MinE in the tip frame, while on the right of Eqn.~\ref{EQN:unscaledF} is a source term due to MinE release at the depolymerizing filament tip.  If we rescale all lengths by $L$ (so $\tilde{z} \equiv z/L$) and define dimensionless fields $\tilde{B} \equiv B/B_{max}$ and $\tilde{F} \equiv Da F /(v L)$, then we can consider the scaled steady-state equations:
\begin{eqnarray}
	\tilde{B}' & =  & - \tilde{\sigma}_1 \tilde{F} (1-\tilde{B}) \text{ , for } \tilde{z}>0 \\	\label{EQN:B}
	\tilde{F}'' & = & -\alpha_L \tilde{F}' -\tilde{B}'-\tilde{B}(0) \delta(\tilde{z}).	\label{EQN:basicF}
\end{eqnarray}
The boundary conditions are $\tilde{F}(\pm 1)=0$ and $\tilde{B}(1)=\rho_0$.  The behavior is controlled by the dimensionless parameters $\tilde{\sigma}_1 \equiv \sigma_1 L^2/(Da)$ and $\alpha_L = v L/D$, as well as by $\rho_0$. We integrate Eqn.~\ref{EQN:basicF} for $ \tilde{z}<0$ where $\tilde{B}=0$, and impose flux conservation of MinE at the boundaries with 
$\tilde{F}'(-1)-\tilde{F}'(1)= \rho_0$. For $\tilde{z}>0$ the equations are then integrated numerically to find the steady-state.  

Following the discussion of the stochastic 3D implementation, we expect $\alpha_L$ to be small, and anticipate that it is irrelevant for the E-ring structure -- leaving only $\rho_0$ and $\tilde{\sigma}_1$ as relevant control parameters. Nevertheless, the 1D treatment allows us to explore this assumption. We find 
that $\alpha \lesssim 0.05$ does not change the observed E-ring steady-state structure by eye, while we expect $\alpha_L \approx 0.01$ at room temperature {\em in vivo} --- and even lower values for weak E-rings. The four-fold speedup observed for the Min oscillation at body temperature \cite{Touhami2006} puts the depolymerization speed (i.e. $\alpha$) closer to, but still under, relevance with respect to the structure of the steady-state MinE ring. 

\section{Results}
We can compare results of our 1D deterministic model with our 3D stochastic model using $F \equiv \pi R^2 \rho_{3,av}$, where $\rho_{3,av}$ is the bulk density averaged over the bacterial cross-section. Then the 1D and 3D binding rates of MinE are related by $\sigma_{1}=\sigma_{3}f/(\pi R^2)$, where $f \equiv \rho_{3,local}/\rho_{3,av}$.   We expect that $f$ will vary with distance from the filament tip due to local release at the tip followed by diffusion and capture. We find $f \lesssim 1$ away from the filament tip due to rebinding to the MinD filament, and we expect $f \gtrsim 1$ 
at the filament tip due to local release from the depolymerizing tip.   Effects of multiple filaments ($n$) and filament pitch ($p$) can be included in the 1D model by using the MinD monomer spacing projected along the bacterial axis 
$a$, where 
\begin{equation}
	a=a_0/ (n \sqrt{1+4\pi^2R^2/p^2}).  
\end{equation}	
Differences between the two approaches are either due to the 1D vs. 3D geometry or due to the deterministic vs. stochastic nature of the models.

\subsection{Strong and weak E-rings}
Fig.~\ref{fig:profiles} illustrates the fractional occupation $\rho$ (equivalent to $\tilde{B}$ in the 1D model) of MinE binding sites on the MinD filament vs. distance $z$ along the bacterial axis.   Occupation monotonically decreases from the tip value, $\rho_{tip}\equiv \rho(0)$, due to local rebinding of MinE following depolymerization from the tip. Following the quantification of Shih {\em et al} \cite{Shih2002},  there are a few thousand MinD monomers within a typical bacteria. With $L=2\mu$m and $a_0=5$nm \cite{Hu2002,Suefuji2002}, they can be arranged either in one single helical filament (with $p \approx 0.45\mu$m \cite{Shih2003}) or about 7 straight filaments (with $p=\infty$). In either case, we find (A) a ``strong'' E-ring ($n=1$ shown) with $\rho_{tip} \approx 1$ and a plateau shape of the density profile near the tip. With only one straight filament (B),  we find a ``weak'' E-ring with enhanced density at the tip but no saturation ($\rho_{tip}<1$) and no plateau.  Strong or weak E-rings have, respectively, negative ($\rho''(0)<0$) or positive ($\rho''(0)>0$) curvature at the tip. 

The 1D model profiles %
reasonably match the 3D results away from the filament tips, using $f=0.06$. This best value of $f$ depends on $r_{bind}$. Using the same $f$ near the tips, the 1D model systematically underestimates the fractional occupation. This implies that a larger $f \equiv \rho_{3,local}/\rho_{3,av}$ is appropriate there, in agreement with the increased likelihood that MinE will be found near the tip shortly after it is released at the tip. 

\begin{figure} \begin{center} \includegraphics[width=\linewidth]{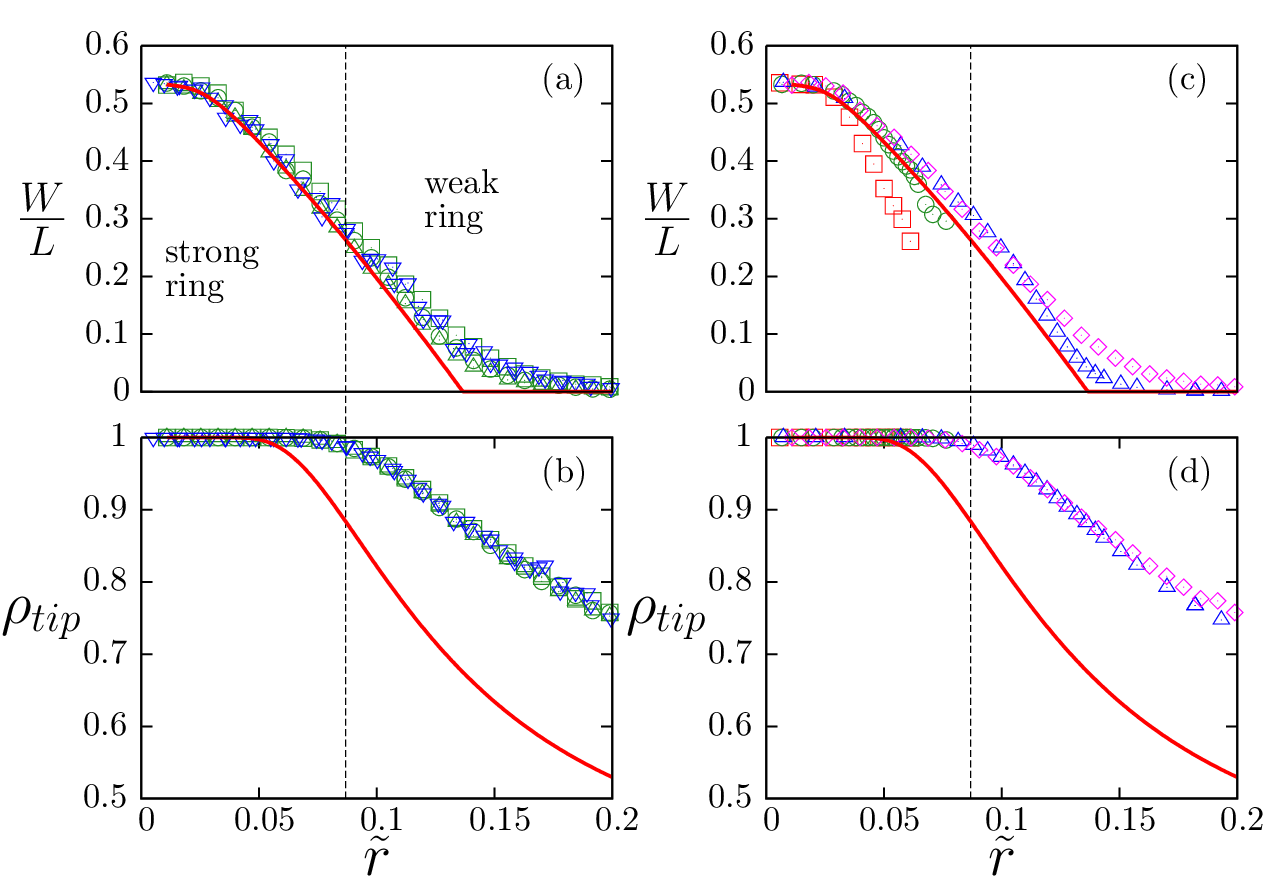}
\caption{MinE profile, characterized by $W/L$ and $\rho_{tip}$, as a function of the scaled aspect ratio $\tilde{r}$. (a, b): Straight filaments ($p = \infty$, $\rho_0=0.35$). 1D model results are shown with solid lines (using $f=0.06$); 3D stochastic results are indicated by symbols. Single filaments ($n=1$, green data) for $L=1~\mu$m ($\square$), $L=2~\mu$m ($\circ$) and $L=3~\mu$m ($\triangle$); multiple filaments ($n=$2, or 5, blue data) for $L=1\mu$m ($\triangledown$). (c, d): Helical filaments ($n=1$, $\rho_0=0.35$) with $L/p=$20 ($\square$, red), 10 ($\circ$, green), 4 ($\triangle$, blue) and  0 ($\diamond$, pink). For all these data, $\sigma_3=0.3 \mu m^3/s$, and $R$ is varied to explore $\tilde{r}$. 
Similar results are obtained when $\sigma_3$ is varied.}
\label{fig:2regimes}   
\end{center} \end{figure}

\subsection{Scaling collapse of E-ring width}
For both strong and weak E-rings, we can define the width $W$ of the E-ring such that $\rho(W)=(1+\rho_0)/2$.  Motivated by the importance of the scaled MinE rebinding rate $\tilde{\sigma_1}$ in the 1D deterministic equations and by the correspondence of $\sigma_1$ and $\sigma_3$, we investigated the influence of the scaled aspect ratio $\tilde{r} \equiv \sqrt{f/\tilde{\sigma_1}}=R/L \sqrt{\pi D a / \sigma_3}$  on the profile shape, as characterized by $W/L$ and by $\rho_{tip}$,  in Fig.~\ref{fig:2regimes} for both the 3D stochastic model (symbols) and the 1D deterministic model (lines).  Two regimes are demarcated by a vertical dashed line:  for small $\tilde{r}$  we have a strong E-ring with $\rho''(0)<0$, a saturated tip ($\rho_{tip} \approx 1$), and good agreement between the 1D and 3D models for the E-ring width; for larger $\tilde{r}$ we have a weak E-ring with $\rho''(0)>0$, $\rho_{tip}$ no longer saturated, and a smaller width $W$.  

The agreement between the 3D and 1D results for $\rho_{tip}$ and $W$ at small  $\tilde{r}$ shows that the essential physics of strong E-rings is one-dimensional.  For small enough $R$ the bacterial cross-section is well explored by MinE by the time it has diffused to free binding sites a distance $W$ from the filament tip.  However, by effectively averaging the radial profile the 1D model systematically underestimates the occupation fraction near the tip, as seen with $\rho_{tip}$ in Fig.~\ref{fig:2regimes} and also in the profiles shown in Fig.~\ref{fig:profiles}. The disagreement becomes stronger as $\tilde{r}$ increases, reflecting the increasingly 3D character of the stochastic system at larger aspect ratios.  However, the system still exhibit a remarkable collapse for all values of $\tilde{r}$. This shows that although the 1D model misses important details about the tip enhancement, the scaling behaviour of the 3D system with straight filaments is similar to the 1D model.

As shown in Fig.~\ref{fig:2regimes}(c, d), $\tilde{r}$  also captures the effects of helical MinD filaments.  Smaller pitches lead to stronger E-rings. However, the 3D stochastic results do not show scaling collapse with respect to $\tilde{r}$ as the monomer spacing along the filament $a_0$ is a relevant length-scale in addition to the 
projected axial monomer spacing $a$. Since the 1D model only uses the effective $a$, it incorrectly exhibits perfect scaling collapse.

\begin{figure} \begin{center} \includegraphics[width=\linewidth]{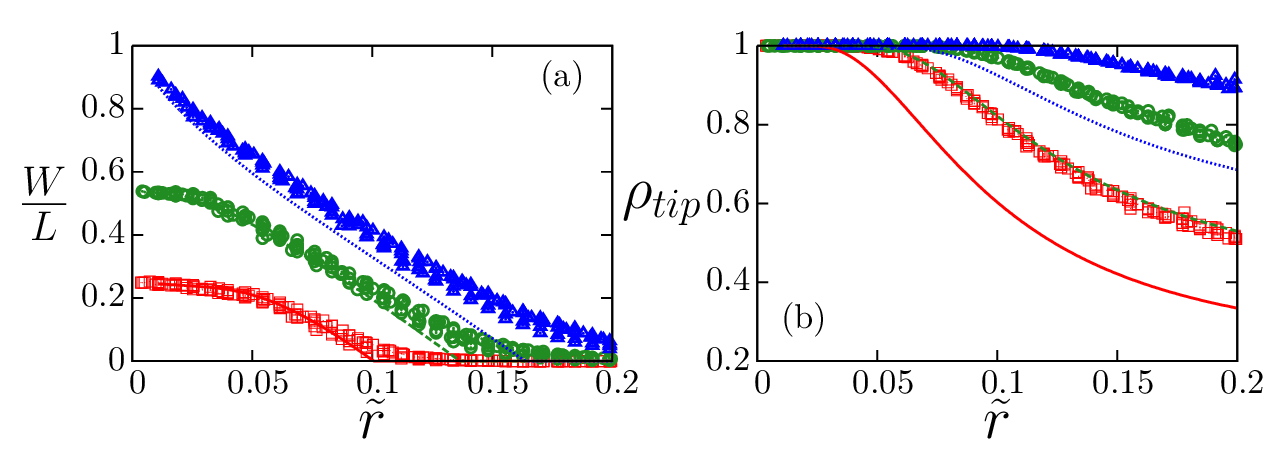}
\caption{(a, b): MinE profile, as characterized by $W/L$ and $\rho_{tip}$ obtained by the 3D model (points) and the 1D model (lines, using $f=0.06$), as a function of $\tilde{r}$ for different values of stoichiometry; $\rho_0=0.2$ ($\square$, continuous lines), 0.35 ($\circ$, dashed lines), 0.50 ($\triangle$, dotted lines). For each stoichiometry, the same collapse as Fig.~\ref{fig:2regimes}(a,b) is obtained: data are compiled for $n$=1, 2, 3, 4 and 5, $L$=1, 2 and 3$\mu$m, 
$p = \infty$, $\sigma_3=0.3 \mu m^3/s$, and $R$ varies to explore $\tilde{r}$.  Similar results are obtained when $\sigma_3$ is varied. }
\label{fig:family} 
\end{center} 
\end{figure}

As shown in Fig.~\ref{fig:family}(a) and (b), $\rho_0$ (the ratio of the number of MinE and MinD particles) also controls the scaling curves of $W/L$ or $\rho_{tip}$ vs. $\tilde{r}$. Agreement between 1D and 3D models for small $\tilde{r}$ and scaling collapse are preserved for each $\rho_0$. 

\subsection{Correspondence with {\em in vivo} Min oscillations}
Experimentally, $W/L \approx 0.3$ is observed in rod-shaped cells \cite{Raskin1997,Hale2001,Fu2001}, where we take $L$ as half the bacterial length.  Using $\rho_0 \approx 0.35$, which is consistent with the ratio of MinE to MinD if we assume MinE are always dimerized \cite{Shih2002}, then from Fig.~\ref{fig:family}(a) we see that $W/L \approx 0.3$ is recovered for $\tilde{r} \approx 0.07$ --- which corresponds to $\sigma_3 \approx 0.3 \mu m^3/s$.  (This $\sigma_3$ is of the same order of magnitude as used in a number of previous models in 3D \cite{Huang2003,Pavin2006} and in 1D \cite{Kruse2002,Drew2005,Tostevin2006} if we assume $R=0.5 \mu m$.)    Interestingly, this indicates that the E-ring of the normal wild-type (WT) Min oscillations is a strong E-ring (with a plateau of MinE occupation near the MinD filament tip) but near the margin between weak (with $\rho_{tip}<1$) and strong. This implies (see Eqn.~\ref{EQN:deltat} below) that the tip occupation $\rho_{tip}$, and hence the depolymerization speed and the oscillation period, will strongly depend on the stoichiometry of MinE to MinD.  Since $k_S/k_I \gg 1$, changes to $\rho_{tip}$ even at the percent level should be significant. Indeed, MinD overexpression leads to a 2.5-fold increase in the period \cite{Raskin1999}.   This also implies from Fig.~\ref{fig:family}(a) that the width of the E-ring will strongly depend on the stoichiometry --- though this has not (yet) been explored experimentally. At a fixed stoichiometry of MinE to MinD ($\rho_0$), we expect that overexpression of Min will increase the number of filaments and/or decrease the pitch. As a result, we expect a slightly stronger E-ring, and a slightly faster period -- as seen \cite{Raskin1999}. 

Optically reconstructed E-rings \cite{Shih2003} show a plateau-like decoration along the MinD filament, consistent with a strong E-ring.  E-rings have also been seen in long filamentous cells \cite{Raskin1997,Hale2001,Fu2001} and exhibit approximately the same width $W$, though both the spacing between MinD caps and the cell length %
are considerably longer in filamentous cells than in rod-shaped cells.  This indicates that the effective $L$ may not be determined by cell shape, but rather by other processes  preserved between rod-shaped and filamentous bacteria such as the length of the MinD filaments or spontaneous lateral release (without MinD hydrolysis) of MinE away from the tip of the MinD filament.

Shih {\em et al.} \cite{Shih2002} identified MinE point-mutants (MinE$^{\text D45A}$ and MinE$^{\text V49A}$) that led to fainter E-rings, and double mutants (MinE$^{\text D45A/V49A}$) that resulted in most of the MinE being cytoplasmic with no strong E-rings. Assembly and disassembly of MinD polar zones continued with no more than doubled periods \cite{Shih2002} --- too rapid to be explained by intrinsic depolymerization alone (in contrast, see \cite{Cytrynbaum2007}). From Eqn.~\ref{EQN:deltat} 
(below) the observed disassembly rates would only require a moderately enhanced $\rho_{tip} \approx 0.9$, i.e. a weak E-ring. 
Indeed, in all of these constructs there appears to be enhanced co-localization of MinE with the MinD polar zones \cite{Shih2002}.   We believe that the lack of visible E-rings in these mutants can be explained with decreased $\sigma_3$ (as suggested previously by \cite{Huang2003}) and/or enhanced spontaneous MinE unbinding away from filament tips.  Local rebinding of MinE near filament tips would still lead to an enhanced $\rho_{tip}$. We predict that the oscillation period in these mutants should be strongly susceptible to the MinE to MinD stoichiometry. 

\section{Transients}
We may use our models to check that the transients before steady-state are fast enough in the context of the normal Min oscillation.  If we initially decorate the MinD filament with  MinE monomers released from $z=-L$ consistent with MinE released from a different depolymerizing MinD cap, we find (data not shown) an initial decoration pattern that has a plateau-like strong E-ring from the beginning (as previously noted \cite{Huang2004}), so that we expect rapid E-ring formation without appreciable delay during Min oscillations (as also observed experimentally\cite{Hale2001,Fu2001}). 

\subsection{Transients before the steady-state {\em in vitro}}
While delays are not observed for E-ring formation during Min oscillations {\em in vivo},  significant delays are observed {\em in vitro}.  MinD binds to phospholipid vesicles in the presence of ATP and undergoes self-assembly, constricting the vesicles into tubes with diameters on the order of 100 nm \cite{Hu2002}. Electron-microscopy revealed that MinD assembles into a tightly wound helix on the surface of these tubulated vesicles with a pitch (helical repeat distance) of only 5 nm. Hu {\em et al.} \cite{Hu2002} report a significant delay (several minutes) for stimulated ATPase activity when small concentrations of MinE were added, while this delay vanished for larger MinE concentrations. Similar delays were seen {\em in vitro} by Suefuji {\em et al.} \cite{Suefuji2002}. Furthermore, the eventual steady-state ATPase activity was smaller for smaller concentrations of MinE \cite{Hu2002,Suefuji2002}.  This has led to the hypothesis of explicit cooperativity of MinE binding, which has then been explicitly included in reaction-diffusion models\cite{Meinhardt2001,Loose2008} and in MinE polymerization in models with MinD polymers \cite{Drew2005,Cytrynbaum2007}.  Here we show that our stochastic model for the MinE ring, with no explicit MinE cooperativity, can recover the MinE concentration dependent ATPase delays and activities observed {\em in vivo}. We conclude that 
MinE cooperativity is not needed to explain the {\em in vitro} results, apart from cooperative effects that arise implicitly from the self-organization of the MinE ring.

We use an ``inside-out'' open geometry corresponding to what is reported {\em in vitro} \cite{Hu2002}, with a narrow phospholipid cylinder that is tightly wound by MinD filaments.  MinE, when released by a depolymerizing filament tip, will diffuse {\em outside} the cylinder. We consider a helical MinD filament of radius $R=50$ nm and pitch 5 nm (equal to $a_0$). Upon MinD depolymerization, we allow any released MinE to diffuse until either it binds to an available MinD binding site or it is absorbed by the boundaries at $z=\pm L$. We impose reflecting boundary conditions at $r=R$, but otherwise allow  MinE to diffuse freely for $r>R$.  Our stochastic 3D model is otherwise the same as before though with an emphasis on the transients approaching steady-state. 

\begin{figure}[h]
 \begin{center} \includegraphics[width=\linewidth]{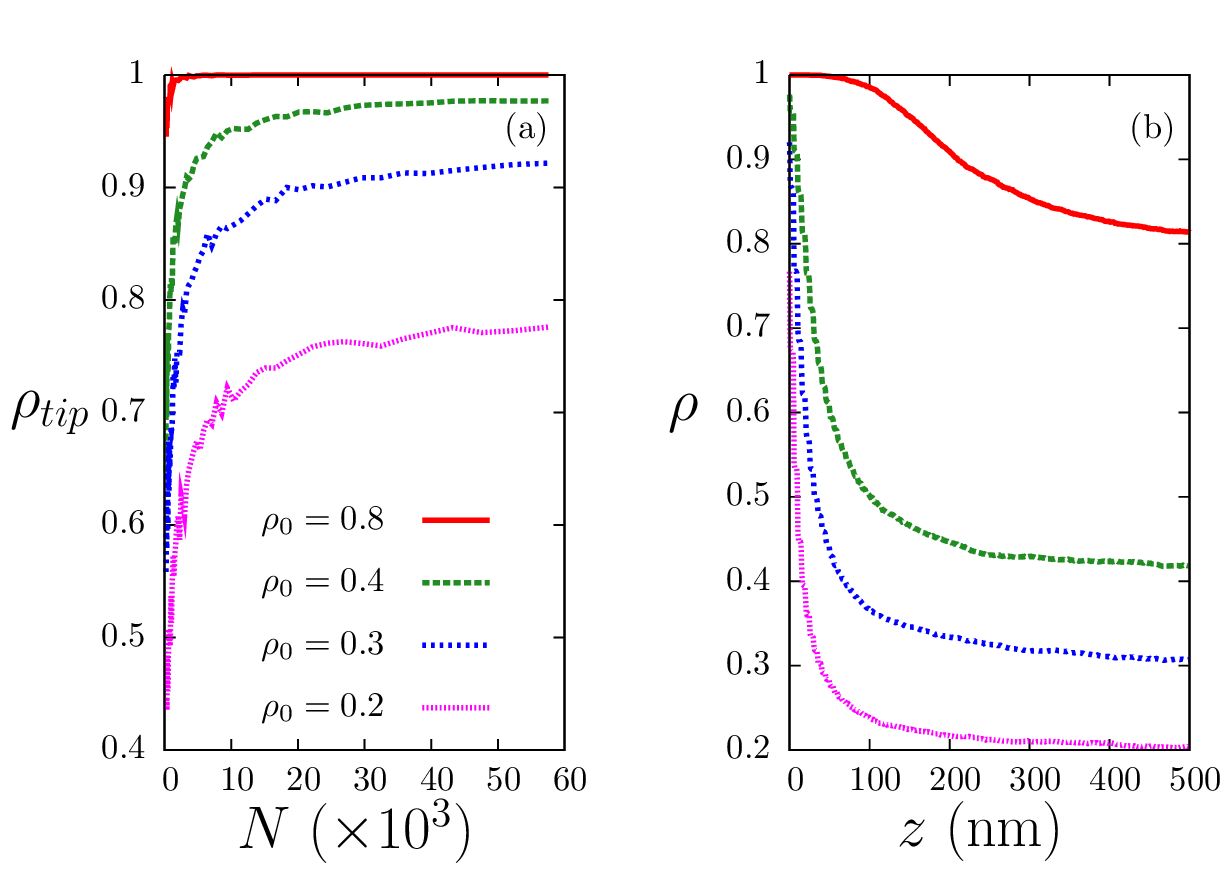}
\caption{Transients and E-ring structure for an ``inside-out'' open geometry appropriate for {\em in vitro} experiments, where a MinD filament is tightly wound on the outside of a cylinder of small radius ($R=50$ nm) with open boundaries at $R=\infty$. (a) Evolution of $\rho_{tip}$ as a function of the number of depolymerization steps $N$ (measured in thousands) after the 
uniform intial conditions for $\rho_0$ equal to $0.8$ 
(solid, red), $0.4$ (long dash, green), $0.3$ (short dash, blue), and $0.2$ (dotted, pink); (b) steady-state $\rho(z)$ as a function of axial distance $z$ along the helical axis for the same $\rho_0$.}
\label{fig:40bis} 
\end{center} 
\end{figure}

The transient to steady-state is shown in Fig.~\ref{fig:40bis}(a), with the fractional occupation of MinE at the MinD filament tip ($\rho_{tip}$) shown as a function of the number of depolymerized monomers from the filament tip, $N$.  The MinE occupation fraction at the MinD filament tip is 
experimentally observable through the ATPase activity (i.e. the MinD depolymerization rate). The initial condition is a uniform occupation $\rho_0$, corresponding to an initially random binding of MinE on the MinD filament.   The larger $\rho_0$ is, the shorter the transient and the stronger the eventual steady-state $\rho_{tip}$. 
Significant enhancement of $\rho_{tip}$ is obtained even for small fractions of MinE. 
For $\rho_0 \gtrsim 0.2$ we see that $\rho_{tip}>0.8$, though, as shown in Fig.~\ref{fig:40bis}(b), 
strong E-rings are predicted only for very large stoichiometry 
($\rho_0 \gtrsim 0.8$). The inside-out {\em in vitro} geometry includes some small radius features (the helical winding of the MinD filament) and some large radius features (no closed boundary at large $r$). The tight helical winding of the MinD filament contributes to long transients, while the semi-infinite radial geometry contributes to the weak E-ring for small and moderate $\rho_0$. 

\begin{figure}[h] \begin{center} \includegraphics[width=\linewidth]{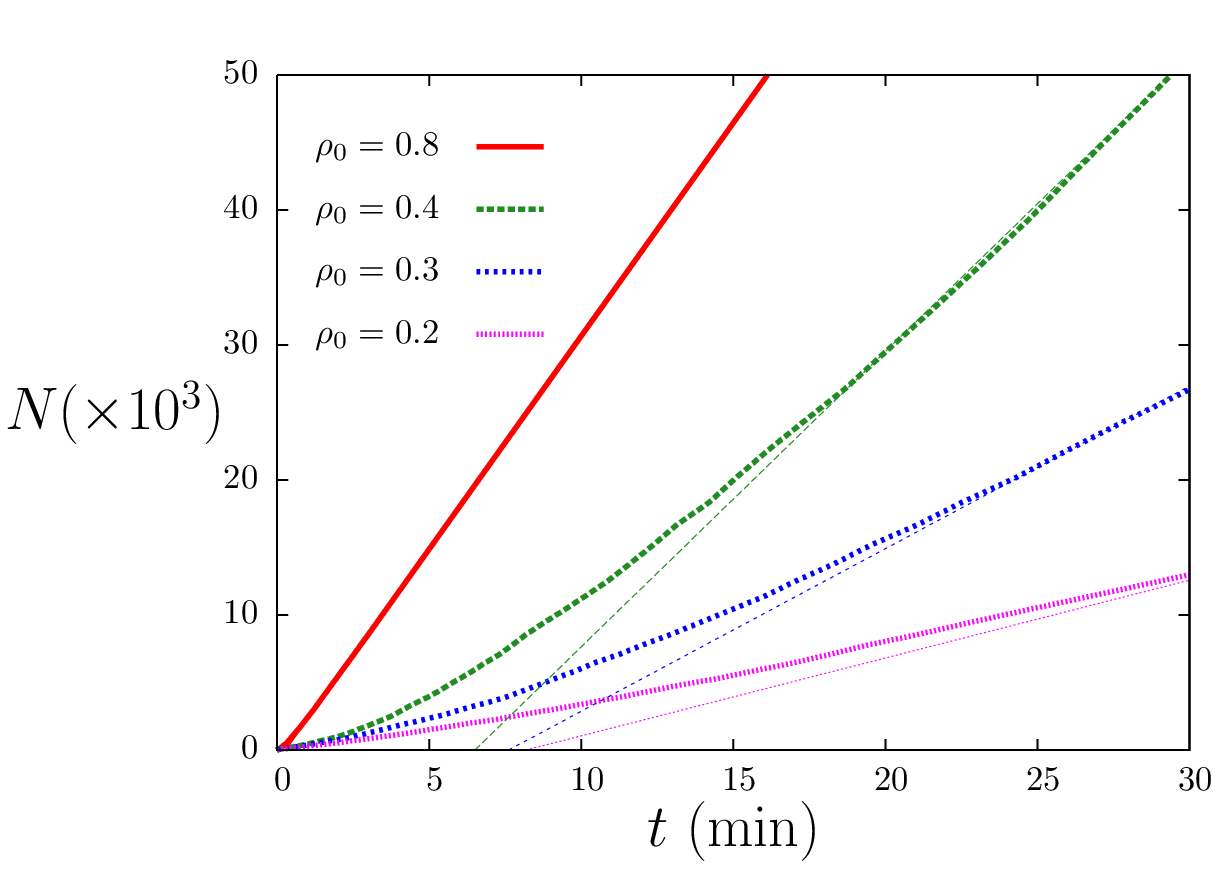}
\caption{For the same inside-out {\em in vitro} geometry described in the previous figure.  Cumulative ATP-ase activity $N(t)$ (measured in thousands of depolymerization steps ) versus time $t$ for various $\rho_0$. Asymptotic behavior are plotted as thin dotted lines.}
\label{fig:invitro} 
\end{center} 
\end{figure}

To convert the number of depolymerization steps $N$ to a time $t(N)$ we need to sum the average time for each step, which will depend on $\rho_{tip}$: $t(N)=\sum_{n=1}^{N}\Delta t(n)$ where,  
\begin{equation}
 \Delta t(n)= \rho_{tip}(n)/k_S+ (1-\rho_{tip}(n))/ k_I.
	\label{EQN:deltat}
\end{equation}
The timesteps are determined by $k_I$ when the tip of the MinD polymer is unoccupied by MinE and $k_S$ when it is occupied.  Using  $k_S/k_I =20$ \cite{Suefuji2002} and  $k_S = 1/(20 ms)$ given by the maximal depolymerization speed {\em in vivo} (assuming $\rho_{tip} \approx 1$, with a strong E-ring)
\cite{depoly},
we plot the cumulative total ATPase activity $N(t)$ (equal to the number of depolymerization steps) vs. elapsed time $t$ in Fig.~\ref{fig:invitro}.
 
The stoichiometric ratio of MinE to MinD corresponds to $\rho_0$ if the MinE mostly binds to available MinD before depolymerization proceeds significantly.  For small amount of MinE (typically $\rho_0 \lesssim 0.3$) we obtain a significant delay of about 5 minutes, corresponding to the ATPase delay seen {\em in vitro} \cite{Hu2002,Suefuji2002} ; and for larger MinE amounts ($\rho_0$ going to 1) the delays decrease towards zero also in agreement with {\em in vitro} studies.  When the steady-state $\rho_{tip}$ is reached, the ATPase rate will also be in a steady-state as indicated by the linear asymptotes in Fig.~\ref{fig:invitro}. Since $\rho_{tip}$ can be large even for smaller $\rho_0$, we expect the ATPase rates to be comparable for moderate or larger $\rho_0$, as seen {\em in vitro} \cite{Hu2002,Suefuji2002}. For smaller $\rho_0$ the steady-state ATPase activity is reduced, as also observed.  

We conclude that the delay of ATPase activity seen {\em in vitro} 
is determined by the time needed to reach the steady-state $\rho_{tip}$. We see that  it is considerably longer in an open than in a closed geometry. 
Our local rebinding model recovers the delays seen {\em in vitro} without any explicit MinE cooperativity (see, conversely, \cite{Meinhardt2001,Hu2002,Suefuji2002,Drew2005,Cytrynbaum2007,Loose2008}).  

\subsection{E-ring instability}
In the tip-frame, the MinD filament tip is bistable during Min oscillations \cite{Cytrynbaum2007} and the formation of the E-ring switches the filament tip between polymerization and depolymerization.  While long transients for this switching are not expected during Min oscillations {\em in vivo} because of initially non-uniform tip decoration \cite{Huang2004}, we may ask about the transient to form the E-ring from a non-oscillating state --- such as seen experimentally after exposure to high levels of extracellular cations \cite{Jericho2009}.  We consider a MinD filament that is initially uniformly decorated with MinE. To tractably include the MinD polymerization dynamics, we use a uniform (mean-field) bulk MinD density $\rho_D$. Because we are interested in the initial slow stages of E-ring formation, we consider MinE binding only near the tip with occupation fraction $\rho_{tip}$ (initially equal to $\rho_0$)

The net polymerization rate of a MinD filament is $R \equiv k_+ \rho_D - (\rho_{tip} k_S + (1-\rho_{tip}) k_I)$, where $\rho_D$ is the bulk MinD monomer concentration and $k_+$ controls MinD monomer addition.  Depolymerization of $n$ monomers from a single tip will enhance $\rho_{tip}$ due to local rebinding of MinE, so that $dR/dn = k_+ /V-(k_S-k_I) d\rho_{tip}/dn$ for cell volume $V$.  The depolymerization time per monomer is $\Delta t \approx 1/k_I$ for an initially weak E-ring (with $\rho_{tip}$ small), and the change in tip occupation 
in one depolymerization step will be proportional to both the number of MinE released ($\rho_{tip}$) and the 
locally available binding sites ($1-\rho_{tip}$), so that
 \begin{equation}
 	\frac{dR}{dt} = k_+ k_I/V -A (k_S-k_I) k_I  \rho_{tip}(1-\rho_{tip}),
		\label{EQN:instability}
 \end{equation}
where the constant $A$ is the fraction of MinE that rebind to available sites at the filament tip.  For  $k_S$ sufficiently greater than  $k_I$ this represents an instability ($dR/dt$ growing more negative with time) that will lead to E-ring formation.  We therefore expect that both a significant difference between intrinsic {\em and} stimulated ATPase activity of MinD and significant intrinsic ATPase activity are needed for E-ring formation, and hence for the initiation of Min oscillations.   

We have neglected any lateral unbinding of MinE from the MinD filament, which will kill the instability if $dR/dt$ is small enough. We also neglect the presence of other MinD filament tips, which will buffer the bulk MinD density and reduce the effect of the $k_+$ term in Eqn.~\ref{EQN:instability}.  These effects will shift the threshold, but will not change the presence of the E-ring instability.  

Since $\rho_{tip} \simeq \rho_0$ initially, we also predict from Eqn.~\ref{EQN:instability} that {\em both} low and high proportions of MinE to MinD will also preclude Min oscillations by making the MinD filament tip initially stable against depolymerization.  However, using $k_+ = 100/ (\mu M s)$ \cite{Cytrynbaum2007}, $A \approx 1$, and $V = 1 \mu m^3$ we estimate a tiny stoichiometry threshold of $0.003$ (for $\rho_0$ or $1-\rho_0$).  While our predicted stoichiometry thresholds are unlikely to be relevant {\em in vivo}, they may be approachable {\em in vitro}. We also note that 
initially slow E-ring formation dynamics
near the instability threshold should be observable when Min oscillations are restarted after being halted \cite{Jericho2009}. 

Previous models of the full Min oscillation have found limiting MinE:MinD 
stoichiometries, either both low and high \cite{Kruse2002,Howard2003,Drew2005} or just high \cite{Huang2003,Tostevin2006}.  Sufficiently low stoichiometries may not have been explored in the later models. 
Conversely, Min oscillations have always been seen {\em in vivo} with moderate stoichiometry changes \cite{Raskin1999}.  It would be desirable for a more systematic exploration of the role of stoichiometry on Min oscillations, given the predicted stoichiometry limits for the existence of oscillations predicted in this and other models. 

 \section{Discussion}
We have presented a model of the self-assembly of the MinE-ring within single {\em E. coli} bacteria, without invoking either MinE cooperativity or MinE polymerization.  We highlight the difference between strong E-rings, with $\rho_{tip} \approx 1$, essentially 1D physics and a maximal depolymerization speed, and weak E-rings with $\rho_{tip}<1$ that have 3D physics with depolymerization speeds that sensitively depend on the parameters, and especially on the amount of MinE in the cell. In contrast to previous filamentous models that had only strong E-rings \cite{Drew2005,Cytrynbaum2007}, our model shows how changing the stoichiometry of MinE and MinD can change the oscillation period through the depolymerization speed of MinD filaments. MinE-rings in non-polymeric reaction-diffusion models \cite{Howard2001,Howard2003,Huang2003,Meinhardt2001,Kruse2002} follow essentially our local rebinding mechanism in the 1D regime, but will deviate from polymeric models for weaker E-rings in the 3D regime where the monomer scale $a_0$ enters. Since the experimentally measured E-ring width indicates that E-rings {\em in vivo} are close to the threshold between weak and strong, the detailed response of the E-ring structure (i.e. the width $W$, or the depolymerization speed via the tip occupation $\rho_{tip}$) to experimental manipulations that change the oscillation period (stoichiometry through $\rho_0$ or, e.g., \cite{Jericho2009}) is unlikely to be correctly captured by 1D or non-filamentous models.  

We have explained the anomalous delays of MinE stimulated MinD ATPase activity seen {\em in vitro} \cite{Hu2002,Suefuji2002}, and have also identified an instability of MinE ring formation that is required to develop from a disordered initial state to the full Min oscillation. We have shown that MinE-ring structure and dynamics can be treated independently of a full Min oscillation model. The instability to E-ring formation, and subsequent MinD filament depolymerization, that we identify neither depends on nor determines the spatial pattern of Min oscillation -- which could be selected by either diffusion and rebinding of MinD \cite{Kulkarni2004} or by phospholipid heterogeneities \cite{Mileykovskaya2005}.  

We have constructed our E-ring model to obtain a steady-state. The steady-state is formed by balancing the MinE entering the system as a bound fraction $\rho_0$ on the MinD filament with the MinD lost by diffusing across the open boundaries at $z \pm L$.  Other geometries, such as an open boundary at $z=-L$ and closed at $z=L$, or a filament tip placed asymmetrically (away from $z=0$), will also lead to a steady-state E-ring that should be qualitatively similar to the one we have described. An extreme example of this is the inside-out geometry we used to describe {\em in vitro} ATPase experiments.  What we have accomplished is to characterize the steady-state, and use it to explore the effects of cell-shape, helical pitch, MinE rebinding rate, and stoichiometry on the E-ring structure.  Our model is expected to be a generic part of full oscillation models that exhibit E-rings.

It is worth speculating on how our simplified E-ring model would be modified by possible additional ingredients within a full model of the Min oscillation.  (1) We do not expect that filament cutting (see e.g. \cite{Tostevin2006,Pavin2006}) will qualitatively affect our results, though it would lead to many more free ends and faster depolymerization. The MinE ring would still only be expected to form near the very end of the MinD filament, and significant depolymerization would only occur within its width $W$ from the end.  Similarly, our results should apply to models without filaments (see e.g. \cite{Meinhardt2001,Kruse2002,Howard2003,Huang2003,Pavin2006}). In that case, we expect that our analytic 1D treatment to be a better approximation due to the absence of an intrinsic monomer spacing $a_0$ that is relevant near the filament tip. (2) We expect that lateral release of bound MinE away from filament tips, without associated cutting, would affect the E-ring profile a distance $\ell = v \tau$ away from the tip (where $v$ is the depolymerization rate, and $\tau^{-1}$ is the lateral release rate). This can be crudely included in our model by placing our boundary conditions at $L \approx \ell$.  (3) We have neglected the rebinding of MinD to the filament tip.  We would expect rebinding to ``poison'' the E-ring by significantly reducing the depolymerization rate -- which would allow further rebinding.  This appears to be observed in the occasional E-ring reversal {\em in vivo} \cite{Hale2001,Shih2002}.  While interesting, poisoning appears to be typically avoided during Min oscillations --- perhaps by filament cutting or by lateral MinE release and re-binding, neither of which have been experimentally characterized --- and so we are justified in neglecting it for steady-state E-rings. Poisoning may however weaken the E-ring instability described by Eqn.~\ref{EQN:instability}, and this deserves further study. The next step is to develop a full 3D Min oscillation model with MinD filaments but without MinE polymerization.

Previous work has considered the steady-states of semi-infinite filaments with tip-directed depolymerization enhanced by bound motors (in this paper, bound MinE) \cite{Klein2005}.  That work used a uniform (mean-field) cytoplasmic motor distribution, and obtained tip-enhanced motor density by a combination of diffusion and directed motion along the filament together with a ``processivity'' retention probability $\bar{p}$ for motors at the depolymerizing tip. In contrast, in our model MinE remains immobile on the filament. [Note that advection ($v$) represents the dragging of MinE along with the MinD filament, not motion with respect to the filament.]  Furthermore, we explicitly consider the cytoplasmic MinE random-walk or diffusion upon release from the filament tip. While this does lead to implicit processivity (local retention of MinE), it also correctly allows for rebinding of MinE away from the filament tip.  This physical modeling of the cytoplasmic MinE allows us to consider, e.g., the 3D vs. 1D cross-over, realistic transients for the inside-out {\em in vitro} geometry, and the E-ring width.  Note that the enhanced local rebinding of MinE to the MinD filament upon release is related to ligand rebinding (see, e.g., \cite{Tauber2005}), and similar dimension and geometry dependent effects are seen there.

 \section*{Acknowledgments}
This work was supported financially by Natural Sciences and Engineering Research Council (NSERC), Canadian Institutes for Health Research (CIHR), and Atlantic Computational Excellence Network (ACENET); computational resources came from ACENET and the Institute for Research in Materials (IRM). We acknowledge useful discussions with Manfred Jericho.


\end{document}